# Blending Mathematical and Physical Negative-ness

Tra Huynh, Eleanor C. Sayre, Kansas State University
Email: trahuynh@ksu.edu, esayre@gmail.com

**Abstract:** Expressing physics problems in the form of a mathematical model is one of the most important stages in the problem-solving process. Particularly in algebraic symbolization, understanding the meanings of signs and being able to manipulate them becomes a challenging task for students, especially when more than two elements in the mathematical expression could carry a negative sign. We use Conceptual Blending theory to investigate how students attribute emergent meaning to the signs and how they articulate different signs in their algebraic symbolization. The data for this research is drawn from oral exams of students enrolled in upper-division physics. The results shed light on students' understanding of algebraic symbols and their competence in formulating and manipulating them.

## Introduction

Mathematics is the common language of science, including physics. By the time university students take intermediate physics, they normally have mastered algebra and practiced calculus in the prerequisite math courses. Accordingly, the mathematical difficulties they usually encounter are appropriately ascribed to the link between mathematical expression and the physical meaning that they are trying to make. Considerable research in physics education research (PER) is devoted to investigating the relationship between math and physics, especially how students use math to express physics concepts and tell physical stories. Conceptual blending (Fouconnier & Turner, 2002) provides a framework where the idea and structure from multiple input domains are blended to generate new meaning, and thus helps us to investigate how students blend mathematics and physics conceptual knowledge. Previous research on math in physics contexts (Bing & Redish, 2007; Hu & Rebello, 2013) argues that students can construct more effective blends through better mapping. Thus, student difficulties might not be because of lacking prerequisite skills but rather because of the inappropriate blending of math, physics knowledge, and the physical scenario at hand. Other work (Gire & Price, 2014) also uses blending theory to show how students make conceptual meaning of electric field vectors. The difficulty occurs when students' represented vector fields in space are also attributed to the clash between input spaces; that is, when an input element (e.g., spatial extent) could represent two meanings simultaneously (distance in coordinate space and magnitude of the field).

In this study, we focus on students in an upper-division Electromagnetism I course with heavy use of mathematics to understand electric and magnetic fields. As a group, their mathematical skills and physics intuitions are substantially more advanced than introductory students, but they still encounter algebraic difficulties with negative signs. The negative sign and its associated concepts in learning and teaching have been studied extensively in k12 contexts back to 1972, but not with upper-division, math-heavy science students. When it comes to electromagnetic topics, the negative sign could be affiliated with different possible elements, such as the charge, vector field, etc. We use conceptual blending theory to account for the meaning that students associate with the sign and to explain the difficulties they face when they manipulate all negative signs at the same time.

## Methods

We collected video data of oral exams done by students in an undergraduate upper-division Electromagnetism I course which enrolls about 20 senior physics students and covers the theory of electric and magnetic fields in vacuum and matter. Our data are drawn from the first oral exam in the fourth week of class, after the class has covered major topics such as Coulomb's law, Gauss' law, and the method of separation of variables.

In an oral exam (Sayre, 2014), students individually work on the board and are encouraged to talk and explain their reasoning to the instructor as they move through the problem. The problem they are exploring is finding the electric field on an axis caused by two equal and opposite charges located on the same axis. The problem appears superficially easy -- it is often given to introductory students -- but requires careful attention to the sign of algebraic expressions and high expectations of consistency among different directions, values, and signs. Four students (all male; three white, one Asian) solve this problem as part of their exam and none of them succeed on their first attempt. The similar struggles across all students become a salient point requiring investigation of students' competence to deal with algebraic signs. We perform moment-by-moment analysis to investigate students' reasoning and construct multiple blends to obtain different emergent meanings that student endow the sign with. In this paper, we introduce a case study on one student whose reasoning is typical of his peers and who eventually arrives successfully at the correct answer.

## Blending between directionality and sign

Conceptual blending theory accounts for how people formulate meaning. Blending, the central action of the theory, is a mental operation of a mental network that generates new meaning. The network consists of at least two input spaces containing information from discrete domains, a generic space containing the common structure and information between input spaces, and a blended space where the new information, whose structure does not exist in any input, emerges. A blending process of generating new meaning has three stages (composition, completion, and elaboration), mostly happening at the subconscious level.

We propose three different blends among the directionality and the algebraic signs, which lead to three different emergent meanings that cover the situation at hand. Because these two input spaces may produce different blends that lead to different conclusions about the physics, we contend that this problem is difficult because selecting the appropriate blend is difficult, not because of an inherent difficulty in choosing input spaces or building each blend. Because vectors have both magnitude and direction, a full analysis of students' blending as they figure out electric fields would require blends for both the magnitude of the vectors and their direction. Observationally, we notice that students tend to treat magnitude as a separate problem as direction, so for brevity we focus only on their directionality reasoning here.

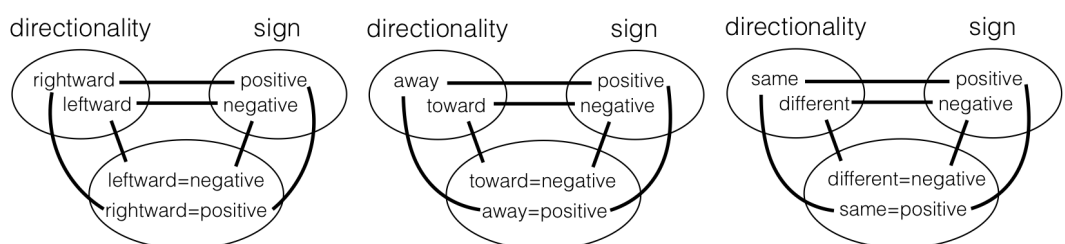

Figure 1. Three blends between directionality and sign. From left to right: for space-fixed coordinates, for body-fixed coordinates, and for comparing the signs of vectors.

In each blend, there are two input spaces which are main characters in our problem, *directionality* and *sign*. The first blend runs when a space-fixed problem is involved, for example a coordinate axis in a one-dimensional problem. The axis could be positive leftward or rightward. From *directionality*, rightward and leftward map to positive and negative (respectively) from *sign*, projecting forward the convention that leftward is negative and rightward is positive. Running the blend yields a vector (e.g. $\vec{x}$) which is positive when it points to the right. This is the usual convention for one-dimensional space-fixed coordinate systems in physics.

Alternately, one could select away and towards from *directionality* to map to positive and negative in *sign* (respectively). This is common in body-fixed coordinate systems: moving away from me is positive velocity and moving towards me is negative; radial vectors are positive away from the source. In the case of the electric field caused by a charge, we add the effect of the sign of the charge into the blend with *directionality* and *sign*. In this further blend (not pictured) a positive charge maps to the positive-away part of the blended *directionality-sign* space, generating a blend where the electric field of a positive charge points away from the charge.

Another significant emergent meaning of the sign comes from the relative direction of two vectors, shown on the right in Figure 1. The characteristics of *directionality* are now "sameness" or "differentness". In a one-dimensional system, "different" and "opposite" are the same, so we kept the label of the more general case. Formally, the association with the sign might come from the mathematical property of the inner product of two vectors; the inner product of two opposite vectors is negative. Apart from comparing the sign of $\hat{x}$, $\hat{E}$, and $\hat{R}$, (the distance vector from the point charge to the field point), the meaning emerged from this comparative blend also sometimes shows up when students consider the interference of two fields. For instance, $\vec{E}_1$ and $\vec{E}_2$ are destructive if their directions are opposite and thus one can insert a negative sign accordingly to account for that destructiveness in the expression for the total field strength. Noting that, the constructiveness and destructiveness are inherently affiliated with the relative direction meaning. Therefore, using these associations requires consistency and care to not double account for this meaning.

## Example and analysis

Oliver, our case study student, uses math explicitly to show exactly what he thinks of each element in the expression. Additionally, he is also good at thinking out loud and expressing his thoughts in detail.

The problem is given to Oliver in both verbal and diagrammatic forms. "Suppose we have two charges at -a and +a [as per the charges in figure 2]. What does the electric field look like along the x-axis?" The problem is typically solved by dividing the whole region into three regions: left of both charges, right of both, and the region between them. The electric field contribution from each charge varies in each region both in direction and magnitude and is commonly drawn like the arrows in figure 2. The total electric field will obey the principle of superposition in all regions.

Table 1: Oliver's solution

| | Time | Mathematical expression | Blend |
|---|---|---|---|
| 1 | 7:56 | $\vec{E} = E_{-q} + E_q = k\frac{-q}{(x+a)^2}\hat{x} + k\frac{q}{(x-a)^2}\hat{x}$ | Blend for body-fixed coordinate, blend of comparing $\hat{x}$ and $\hat{R}$ |
| 2 | 10:53 | $E_1 = k[\frac{-q}{(x+a)^2} - k\frac{q}{(x-a)^2}]\hat{x}$ | Blend of comparing $\hat{E}_q$ and $\hat{E}_{-q}$ (destructiveness) |
| 3 | 12:18 | $E_1 = k[\frac{-q}{(x+a)^2} + \frac{q}{(x-a)^2}]\hat{x}$ | Blend of comparing $\hat{E}_q$ and $\hat{E}_{-q}$ |
| 4 | 13:15 | $E_1 = k[\frac{q}{(x+a)^2} - \frac{q}{(x-a)^2}]\hat{x}$ | Blend of comparing $\hat{E}_q$, $\hat{E}_{-q}$, and $\hat{x}$ |
| 5 | 15:36 | $E_2 = k\frac{-q}{(x+a)^2}(-\hat{x}) + k\frac{q}{(x-a)^2}(-\hat{x})$ | Blend for body-fixed coordinate, blend of comparing $\hat{x}$ and $\hat{R}$ |
| 6 | 17:56 | $E_2 = kq[\frac{1}{(x+a)^2} + \frac{1}{(x-a)^2}](-\hat{x})$ | Blend of comparing $\hat{E}_q$, $\hat{E}_{-q}$, and $\hat{x}$ |
| 7 | 19:30 | $E_3 = k\left[\frac{q}{(x+a)^2}(-\hat{x}) + k\frac{q}{(x-a)^2}\hat{x}\right] = k[\frac{-q}{(x+a)^2} + \frac{q}{(x-a)^2}]\hat{x}$ | Blend of comparing $\hat{E}_q$, $\hat{E}_{-q}$, and $\hat{x}$ |

Oliver starts by writing down the superposition formula of the total electric field: $\vec{E} = E_q + E_{-q}$, and moves to define each contribution using Coulomb's Law, as is appropriate to the problem (Table 1). It takes him a long time to decide what the denominator looks like so that it is consistent with the definition of $\hat{R}$. The formula of the electric field due to a point charge $\vec{E} = k\frac{q}{R^2}\hat{R}$ suggests Oliver should blend the charges' value with the direction of the electric field vector and compare the distance vector $\hat{R}$ to $\hat{x}$. Then, as Oliver thinks that "$\hat{R}$ is along x direction", a positive sign which is commensurate with the "sameness" in $\hat{R}$ and $\hat{x}$ direction is added while he replaces $\hat{R}$ by $\hat{x}$. Oliver concludes that the total electric field along the x axis will be as shown in line 1. We see that Oliver could have arrived at the correct answer if he defined correctly the relationship between $\hat{R}$ and $\hat{x}$ for each charge. However, as vector $\hat{R}$ is not clearly shown on the diagram, this output from the blend of comparing $\hat{R}$ and $\hat{x}$ leads to an array of confusion and conflict with other blends that he uses later.

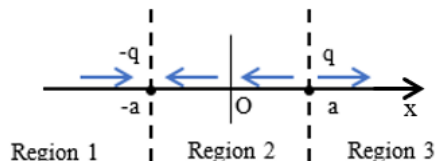

Figure 2. Diagram with the two charges on the x-axis at +/- a. Regions 1-3 and the origin O are marked, and the blue arrows designate the direction of the electric field in each region.

Noticing that the electric field of the positive charge does not always point in the $\hat{x}$ direction as shown in his mathematical expression, Oliver decides to divide the given region into three smaller ones and defines the field vectors on the diagram, as shown in Figure 2. With the diagram, the blend of comparing the vectors, especially $\hat{E}$ and $\hat{x}$, can be run easily and becomes more trustful to Oliver compared to other blends. Later, we observe him trying to fit the sign coming from other blends with this one.

In region 1, Oliver quickly realizes that the contribution of the negative charge to any field point is greater than that of the positive charge. He inserts a negative sign between the two terms accounting for the effect of destructiveness of two component fields as they are in opposite directions. He concludes: "It would be this term [E₋q] minus this term [Eq]" (line 2). However, as he is halfway through recording the expression on the board, Oliver expresses suspicion because both terms are now negative. This result clearly conflicts with their relative direction because he has double associated their opposite direction with inappropriate application of destructiveness. Oliver tries hard to determine where another negative sign could come from, such as the denominator, to cancel one negative sign for the whole term. Finally, he decides to absorb the destructive meaning of the sign into the opposite-direction meaning of the electric field vector and changes the second negative sign of the whole term back into the plus sign (line 3), which supports the fact that they are in opposite directions.

However, Oliver has not considered the sign commensurate with the relative direction of $\hat{E}$ and $\hat{x}$, leading to his solution having the opposite sign to the correct answer.

Reading off the physical sense of Oliver's mathematical expression again, the instructor points out that $E_q$ is pointing in the $\hat{x}$ direction. Oliver starts getting frustrated. He knows that the final solution should have a negative sign in $E_q$ to be consistent with the blend of comparing $\hat{E}$ and $\hat{x}$ that is obviously shown on the diagram. However, Oliver gets stuck manipulating the signs coming from all the sources and ascribes a general meaning to the final sign left in the expression. Eventually, Oliver changes the sign of the terms such that they agree with the emergent meaning of comparative blend between $\hat{E}$ and $\hat{x}$. Oliver confirms his final solution (line 4) and explains his line of reasoning: "Because… see the charges, I should have just figured it out […] which direction it is. This is exactly what is changing the signs, not necessarily the sign of the charge." In other words, Oliver has successfully affiliated the sign's meaning of the effect of charge on the field and the superposition into the sign's meaning of the relative direction among component electric fields and $\hat{x}$.

Moving to the other two regions, we observe Oliver encounter the same struggles with multiple meanings of the sign; however, he's faster at selecting productive blends. Rewriting the mathematical expression in such a way that the minus signs are put next to the elements they belong to (line 5) helps Oliver better distinguish and understand the meaning of the signs. Eventually, deciding to be consistent with the blend of comparing the relative direction of component electric field and $\hat{x}$, Oliver arrives at the correct answers for regions 2 and 3, noting that he should "not worry about the sign [of charge], just worry about the field point [field direction]" (line 6, line 7).

## Discussion and conclusion

As shown in the case study, the problem of two charges appears easy but requires effort and consistency in considering where the signs belong and their meaning. Oliver typically struggles with combining ideas which seems to be easy. Using conceptual blending, we look at the fine-grained structure of the sign and thus, explore his reasoning meticulously, and how he successfully blends the physical and mathematical sign.

We have associated the algebraic sign with three different meanings in this electrostatics problem, which emerges from three blends of directionality with signs. In the cases of electric field with interference, the sign charge input and the constructiveness or destructiveness also play a role in the blend. We argue that the challenging part of this problem is attributed to the complexity of selecting different blends for sign and directionality. The sign could be affiliated with the relationship among $\hat{E}$, q, $\hat{R}$, and $\hat{x}$, or with any of them individually. The student in this case study shows his strong competence in negative number arithmetic, for example, operation of multiplication. However, to successfully deal with algebraic signs, the student has needed to recognize these different blends and productively select among them. In the literature of using conceptual blending theory in science education research, this difficulty could be explained by the clash between different input spaces, where a sign, either positive or negative, could carry multiple meanings in a comparative blend.

The sign is not the only problematic algebraic element in this problem. The other students in our data also struggled with determining the distance from the two point-charges to field point in different regions. However, as we are interested in how students work on the meaning of the negative sign, we purposefully picked Oliver, who had much less confusion about the distance, and focused on his reasoning on charge and directionality.

The problem of introductory level problems when introduced to intermediate-level students helps expose students' understanding of mathematics and their competence in using mathematics to express physical concepts. The salient algebraic complexity of this problem has suggested these types of problems could be an interesting electromagnetism problem in physics education research, leading to greater insight into student algebraic thinking.

## Acknowledgments

This work was partially supported by NSF DUE-1430967.